\documentclass[aps,prl,preprint,superscriptaddress,showpacs,floatfix,longbibliography]{revtex4-1}

\usepackage{mathrsfs}
\usepackage[cmex10]{amsmath}
\usepackage{amsfonts}
\usepackage{graphicx}
\usepackage{textcomp}
\usepackage{color}
\usepackage[caption=false]{subfig}
\usepackage{color}
\usepackage[novolumeabbr]{unitsdef}
\usepackage{esint}
\usepackage{accents}
\usepackage{bbm}

\newcommand{\ce}[1]{{\textrm{#1}}}

\renewcommand{\vec}[1]{{\boldsymbol{#1}}}
\newcommand{\op}[1]{\hat{\boldsymbol{#1}}}

\newcommand{\cross}{\times}

\newcommand{\unitvec}[1]{{\boldsymbol{\mathbbm{#1}}}}
\newcommand{\z}{\unitvec{z}}
\newcommand{\x}{\unitvec{x}}

\newcommand{\vl}{\vec{l}}
\newcommand{\vm}{\vec{m}}
\newcommand{\wex}{\omega_\text{ex}}
\newcommand{\wax}{\omega_e}
\newcommand{\waz}{\omega_h}

\newcommand{\Ms}{M_s}
\newcommand{\muz}{\mu_0}
\newcommand{\wgen}{\omega_\text{gen}}
\newcommand{\jthi}{j^{\text{th}}_1}
\newcommand{\jthe}{j^{\text{th}}_2}

\newcommand{\OU}{
\affiliation{Department of Physics, Oakland University, 146~Library~Drive, Rochester, Michigan,
    48309-4479, USA}
    }

\newunit{\ohmm}{\ohm\unittimes\meter}
\newunit{\THz}{\tera\hertz}
\newunit{\Oe}{Oe}
\newunit{\rad}{rad}

\DeclareMathOperator{\const}{const}

\begin{document}

\title{Antiferromagnetic THz-frequency Josephson-like Oscillator Driven by Spin Current}

\author{Roman Khymyn}
\OU

\author{Ivan Lisenkov}
\email[]{ivan.lisenkov@phystech.edu}
\OU
\affiliation{Kotelnikov Institute of Radio-engineering and Electronics of RAS, 11-7 Mokhovaya street, Moscow, 125009, Russia}

\author{Vasyl Tiberkevich}
\OU

\author{Boris A. Ivanov}
\affiliation{Institute of Magnetism, National Academy of Sciences of Ukraine, Kiev, Ukraine}
\affiliation{National Taras Shevchenko University
of Kiev, 03127 Kiev, Ukraine}

\author{Andrei Slavin}
\OU

\begin{abstract}
The development of compact and tunable room temperature sources of coherent THz-frequency signals would open a way for numerous new applications. The existing approaches to THz-frequency generation based on superconductor Josephson junctions (JJ), free electron lasers, and quantum cascades require cryogenic temperatures or/and complex setups, preventing the miniaturization and wide use of these devices. We demonstrate theoretically that a bi-layer of a heavy metal (Pt) and a bi-axial antiferromagnetic (AFM) dielectric (NiO) can be a source of a coherent THz signal. A spin-current flowing from a DC-current-driven Pt layer and polarized along the hard AFM anisotropy axis excites a non-uniform in time precession of magnetizations sublattices in the AFM, due to the presence of  a weak easy-plane AFM anisotropy.  The frequency of the AFM oscillations varies in the range of 0.1-2.0 THz with the driving current in the Pt layer from $10^8\ampere/\squarecentimeter$ to $10^9\ampere/\squarecentimeter$. The THz-frequency signal from the AFM with the amplitude exceeding 1 V/cm is picked up by the inverse spin-Hall effect in Pt. The operation of a room-temperature AFM THz-frequency oscillator is similar to that of a cryogenic JJ oscillator, with the energy of the easy-plane magnetic anisotropy playing the role of the Josephson energy.
\end{abstract}

\maketitle

An absence of compact and reliable generators and receivers of coherent signals in the frequency range \ilu[0.1--10]{\THz} has been identified as a fundamental physical and technological problem~\cite{bib:Sirtori:2002, bib:Kleiner:2007, bib:Gulyaev:2014}. The existing approaches to THz-frequency generation, including superconductor Josephson junctions (JJ)~\cite{bib:Ozyuzer:2007}, free electron lasers~\cite{bib:Nanni:2015}, and quantum cascades~\cite{bib:Hubers:2010} require complex setups, which limit wide use of these devices. At the same time, it has been demonstrated that ferromagnetic (FM) layered structures driven by a spin-transfer torque (STT) created by a DC spin current~\cite{bib:Slonczewski:1996, bib:Berger:1996}, which compensates magnetic damping, can be used as spin-torque or/and spin-Hall auto-oscillators in the frequency range of \ilu[1--30]{\GHz} ~\cite{bib:Tsoi:2000, bib:Kiselev:2003, bib:Rippard:2004,bib:Demidov:2012, bib:Demidov:2014, bib:Duan:2014, bib:Collet:2016}.

In order to increase the generation frequency it was proposed to use \emph{antiferromagnets} (AFM) rather than FM films as active layers of spintronic auto-oscillators ~\cite{bib:Gomonay:2014,bib:Cheng:2016}. Unfortunately, the traditional method of the STT-induced damping compensation in FM does not work for AFM. To compensate damping in a FM, the DC spin current must be polarized parallel to the direction of the static equilibrium magnetization. However, since AFMs have two magnetic sublattices with opposite magnetizations, the STT decreasing the damping in one of the sublattices increases it in the other sublattice, thus resulting in a zero net effect. Fortunately, the presence in an AFM of two magnetic sublattices coupled by a strong exchange interaction qualitatively changes the magnetization dynamics of AFM~\cite{bib:Ivanov:2014}. In particular, it has been shown, that, in contrast with a FM, the STT acting on an AFM can lead to a dynamic instability in the magnetic sublattice orientation~\cite{bib:Gomonay:2010, bib:Gomonay:2014, bib:Cheng:2015, bib:Gulyaev:2014}, which results in the rotation of the magnetizations of the AFM sublattices in the plane perpendicular to the direction of polarization of the applied spin current~\cite{bib:Cheng:2015,bib:Gomonay:2014,bib:Gomonay:2010}. This mechanism has been already used to experimentally switch the orientation of magnetic sublattices in AFM materials~\cite{bib:Wadley:2016, bib:Kriegner:2016}. However, the STT-induced rotation of the magnetic sublattices in an AFM has not been recognized so far as a possible mechanism of realization of THz-frequency AFM oscillators, since in a magnetically compensated AFM the steady rotation of sublattices does not create any AC spin-current.

In this work we demonstrate theoretically that a simple structure consisting of a metallic layer with a strong spin-orbit interaction (such as Pt) and a layer of a bi-axial antiferromagnetic (AFM) dielectric (such as NiO) can be a base of a tunable room-temperature THz-frequency signal generator.  A DC spin current flowing from a current-driven \ce{Pt} layer and polarized along the \emph{hard} anisotropy axis of the adjacent AFM layer can excite a rotation of the AFM sublattice magnetizations ~\cite{bib:Cheng:2015,bib:Gomonay:2014,bib:Gomonay:2010} that is \emph{non-uniform in time} due to the influence of a weak \emph{easy-plane} AFM anisotropy. This non-uniform rotation results in the THz-frequency spin-pumping back into the Pt layer, creating an AC electric field that can be detected using the inverse spin-Hall effect. The generated signal amplitudes of \ilu[1]{\volt/\cm} in the  frequency range of \ilu[0.1--2.0]{\THz} of the can be achieved for driving DC current densities of $1\times 10^8\ampere/\squarecentimeter$ to $1.1\times 10^9\ampere/\squarecentimeter$, that have been experimentally achieved previously in spin-Hall nano-oscillators ~\cite{bib:Demidov:2012,bib:Demidov:2014,bib:Duan:2014,bib:Collet:2016}.

We also demonstrate, that the equations describing the operation of the  proposed room-temperature AFM-based oscillator are mathematically analogous to the ones describing the oscillators based on superconducting Josephson junctions (JJ) ~\cite{bib:Barone:1982}, with the energy of the easy-plane magnetic anisotropy playing the role of the Josephson energy. Consequently, a number of effects studied previously in JJ oscillators at cryogenic temperatures can also be observed in the room-temperature AFM oscillators. In particular, the inertial nature of the AFM dynamics~\cite{bib:Ivanov:2014} leads to the hysteretic behavior of the AFM oscillator, which, therefore, have two different current thresholds: an ``ignition'' threshold, which is required to start the generation, and a lower ``elimination'' threshold which, in our case, is twice less then the ``ignition'' threshold, defining the minimum current density needed to support the generation, once it has been started.

\section{Results}

\begin{figure}
    \includegraphics[width=0.5\linewidth]{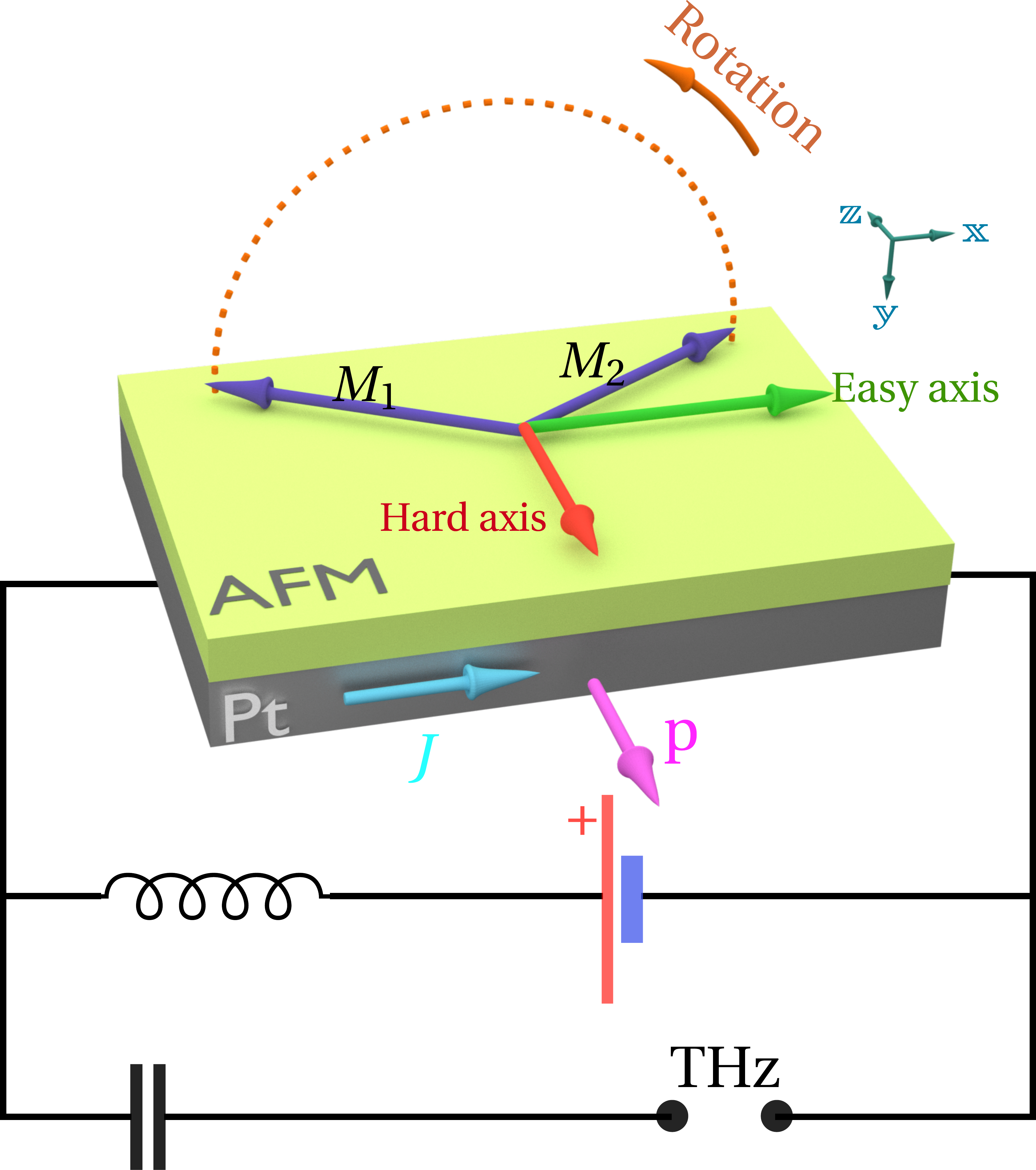}
    \caption{
			Schematic view of the THz-frequency oscillator based on a Pt/AFM bilayer. The AFM hard axis (HA) lies in the bilayer plane perpendicular to the direction of the DC bias current and parallel to the direction of polarization of the spin-current flowing from the Pt layer into the AFM layer $\vec{p}$. Solid dark blue arrows show canted magnetizations under the action of the spin-current. The spin-transfer torque (STT)-induced non-uniform in time rotation of the canted AFM sublattices creates in the Pt layer an AC spin-pumping signal at THz frequencies which is transformed into an AC electric field via the inverse spin-Hall effect in the Pt layer.
		}\label{fig:bi-layer}
\end{figure}

\begin{figure}
    \includegraphics[width=1.0\linewidth]{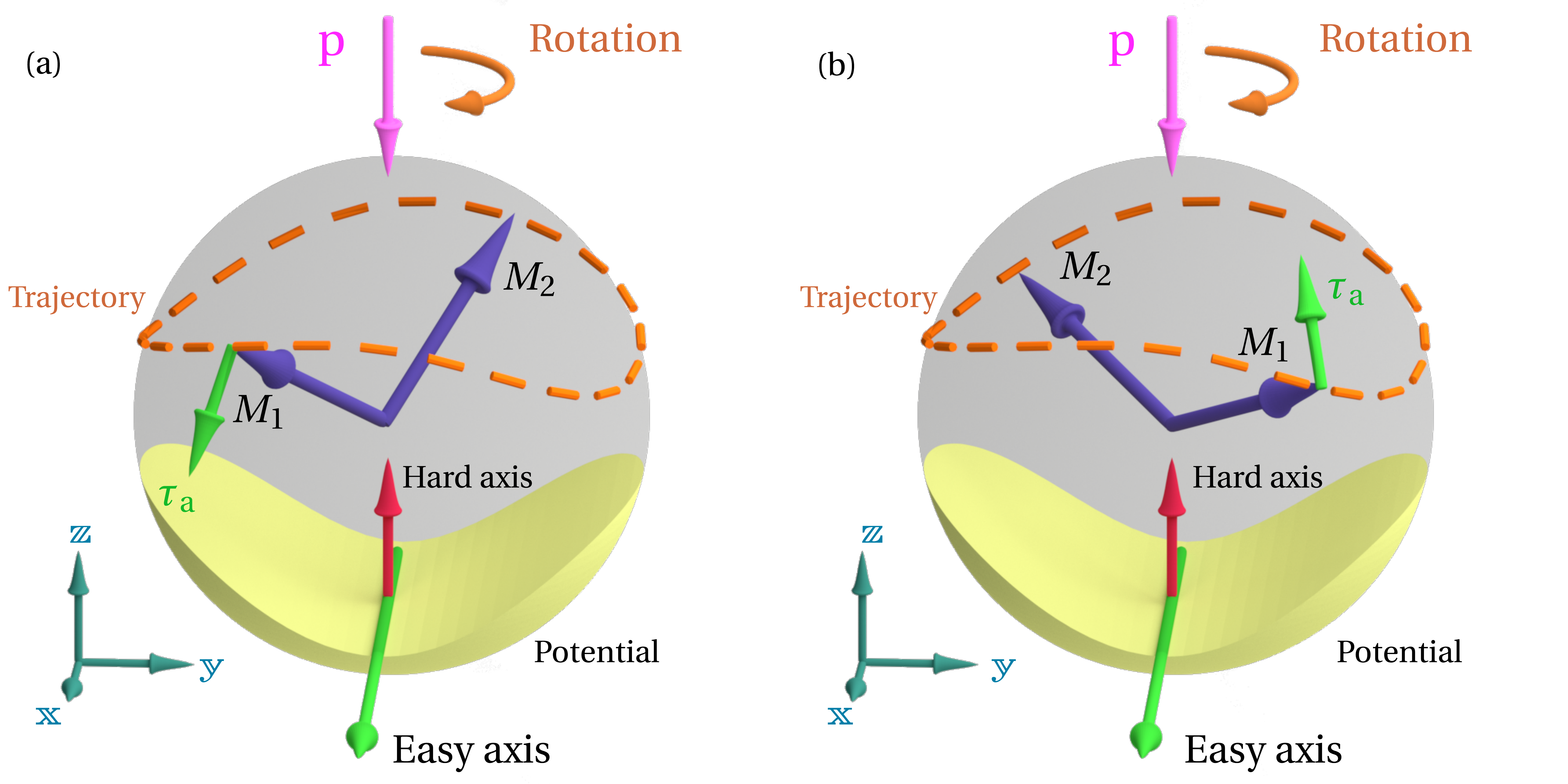}
    \caption{
Schematic representation of the rotating sublattice magnetizations in an anisotropic antiferromagnet under the action of an STT. The presence of the easy-plane magnetic anisotropy (yellow-colored potential) in the AFM layer leads to a variable in time rotation speed of the AFM sublattice magnetizations : (a) in a part of the trajectory the anisotropy torque decreases the tilt angle between the magnetizations, thus decelerating the rotation; (b) in another part of the trajectory the anisotropy increases the tilt and, thus, accelerates the rotation. This nonuniform rotation results in an AC spin-pumping signal in the Pt layer.
		}\label{fig:sphere}
\end{figure}

We consider a bi-layer consisting of a layer of a heavy metal with strong spin-orbital interaction (e.g., Pt) adjacent to a \emph{bi-axial} AFM layer (e.g., NiO), see Fig.~\ref{fig:bi-layer}. A DC electric current passing through the Pt layer creates, via the spin-Hall effect, a perpendicularly-polarized spin current flowing into the AFM layer~\cite{bib:Jungwirth:2016, bib:Moriyama:2015, bib:Daniels:2015}.
Spin current creates a non-conservative spin-transfer torque (STT) on AFM sublattice magnetizations $\vec{M}_j$ ($j=1,2$): $\vec{\tau}_{\text{STT}}=(\tau/\Ms)\vec{M}_j\cross(\vec{M}_j\cross\vec{p})$~\cite{bib:Slonczewski:1996, bib:Gomonay:2010, bib:Linder:2011}, where $\vec{p}$ is the direction of the spin current polarization, $\Ms=|\vec{M}_j|$ is the static magnetization of a sublattice, and $\tau$ is the amplitude of the spin current in the units of frequency~\cite{bib:Tserkovnyak:2005}.
If the spin current is polarized \emph{perpendicularly} to the AFM ground state (along the ``hard'' AFM axis; here we consider only such a configuration), it tilts the magnetizations from their equilibrium opposite orientation $\vec{M}_1 = -\vec{M}_2$, which creates a strong effective field $\vec{H} = H_\text{ex} (\vec{M}_1 + \vec{M}_2)/(2\Ms)$ ($H_\text{ex}\sim10^3\tesla$ is the exchange field) leading to \emph{uniform rotation} (in the absence of the in-plane anisotropy) of the sublattice magnetizations in the plane perpendicular to the spin-current polarization~\cite{bib:Gomonay:2010, bib:Gomonay:2014, bib:Cheng:2015},
see Fig.~\ref{fig:bi-layer}.

The rotation of the tilted sublattice magnetizations in an AFM induces the spin current flowing back, from the AFM to the Pt layer, via the spin-pumping mechanism~\cite{bib:Tserkovnyak:2005,bib:Nakayama:2012,bib:Gomonay:2014}:
\begin{equation}
\vec{j}_s^\text{out}=\frac{\hbar g_r}{8\pi M_s^2}\left(\vec{M}_1\cross \dot{\vec{M}}_1+\vec{M}_2\cross \dot{\vec{M}}_2\right) ,
\label{eq:spin-current}
\end{equation}
where $g_r$ is the spin-mixing conductance.

The exchange interaction is the strongest interaction in AFM, and, even under the action of an STT, the tilt angle of the sublattice magnetizations is small. Then, introducing the AFM Neel vector~\cite{bib:Andreev:1980, bib:Ivanov:2014} $\vl = (\vec{M}_1 - \vec{M}_2)/{2M_s}$ the spin-pumping current~\eqref{eq:spin-current} can be written as:
\begin{equation}
\vec{j}_s^\text{out} \approx \frac{\hbar g_r}{2\pi}\, \vl\cross\dot{\vl} = \z\frac{\hbar g_r}{2\pi} \dot\phi,
\label{eq:spin-current-phi}
\end{equation}
where $\phi$ is the azimuthal angle of the vector $\vl$. Equation~\eqref{eq:spin-current-phi} leads to an important conclusion: \emph{uniform} rotation (with constant angular velocity $\dot\phi =\const$) of the Neel vector creates only a DC spin pumping signal in \ce{Pt}, and creates no AC signal~\cite{bib:Gomonay:2010, bib:Cheng:2015}.

Equation~\eqref{eq:spin-current-phi} is fully analogous to the second Josephson equation connecting voltage bias in a Josephson junction (JJ) with the phase of the supercurrent~\cite{bib:Barone:1982, bib:Savelev:2010}. Similarly to JJ oscillators, to achieve the AC generation we need one more ingredient: a potential ``force'' that that depends on $\phi$. In Josephson junctions this potential comes from the tunneling Hamiltonian (or Josephson energy)~\cite{bib:Barone:1982}. In AFM, the role of the Josephson energy is played by the energy of crystalline magnetic anisotropy $W_a$ \emph{in the easy plane}: $W_a = -\muz\Ms H_e\cos2\phi$, where $H_e$ is the easy-plane anisotropy field.

The presence of the anisotropy leads to a qualitative change in the dynamics of the magnetic sublattices: it creates an additional conservative torque $\vec{\tau}_a(\phi)$, which depends on the orientation of the magnetic sublattices, see Fig.~\ref{fig:sphere}(a,b). Thus the trajectory of each sublattice magnetization is not anymore a planar circle on the sphere's equator~\cite{bib:Cheng:2016,bib:Gomonay:2014}, but is a more complicated curve (see Fig.~\ref{fig:sphere}(a,b)). In one part of the trajectory the anisotropy decreases the tilt angle between the magnetizations $\vec{M}_1$ and $\vec{M}_2$, thus \emph{decelerating} the rotation~(Fig. \ref{fig:sphere}(a)), while in another part the anisotropy increases the tilt and, therefore, \emph{accelerates} the rotation~(Fig.~\ref{fig:sphere}(b)).

At the same time, the trajectory of the Neel vector $\vl$ remains a circle even in the presence of easy-plane anisotropy (see Methods for the derivation details), and can be described with one scalar equation for the angle $\phi$~\cite{bib:Gomonay:2010,bib:Cheng:2014,bib:Cheng:2016}:
\begin{equation}
\dfrac{1}{\wex}\,\ddot\phi + \alpha\,\dot\phi + \dfrac{\wax}{2}\,\sin 2\phi + \sigma j = 0,
\label{eq:phi}
\end{equation}
where $\wax = \gamma H_e$, $\gamma$ is the gyromagnetic ratio, $\alpha$ is the effective Gilbert damping parameter, $j$ is the electric current density in the \ce{Pt} layer, and $\sigma = \tau/j$ is the torque-current proportionally coefficient~\cite{bib:Gomonay:2010, bib:Gomonay:2014}.
This equation is, essentially, a condition of the balance between four torques acting on the sublattice magnetizations: the exchange torque, created by the exchange field, the Gilbert damping torque, the anisotropy torque, and the STT.

Equation~\eqref{eq:phi} is a well-known equation describing dynamics of a massive particle in a tilted ``washboard'' potential. It also \emph{exactly} coincides with the equation describing the superconducting phase in a resistively and capacitively shunted JJ under a current bias~\cite{bib:Stewart:1968}. Here, the role of the energy stored in a capacitor is played by the exchange energy, which accumulates the kinetic energy in the system, and the role of resistance is played by the Gilbert damping. For sufficiently large currents, equation (\ref{eq:phi}) does not have stationary solutions, meaning that the AFM magnetization sublattices lose their stability, and the Neel vector starts to rotate. The electric current density needed to overcome this ``ignition'' threshold is proportional to the easy-plane anisotropy:
\begin{equation}
    \jthi = \dfrac{\wax}{2\sigma}.
    \label{eq:jthi}
\end{equation}

\begin{figure}
    \includegraphics[width=0.7\linewidth]{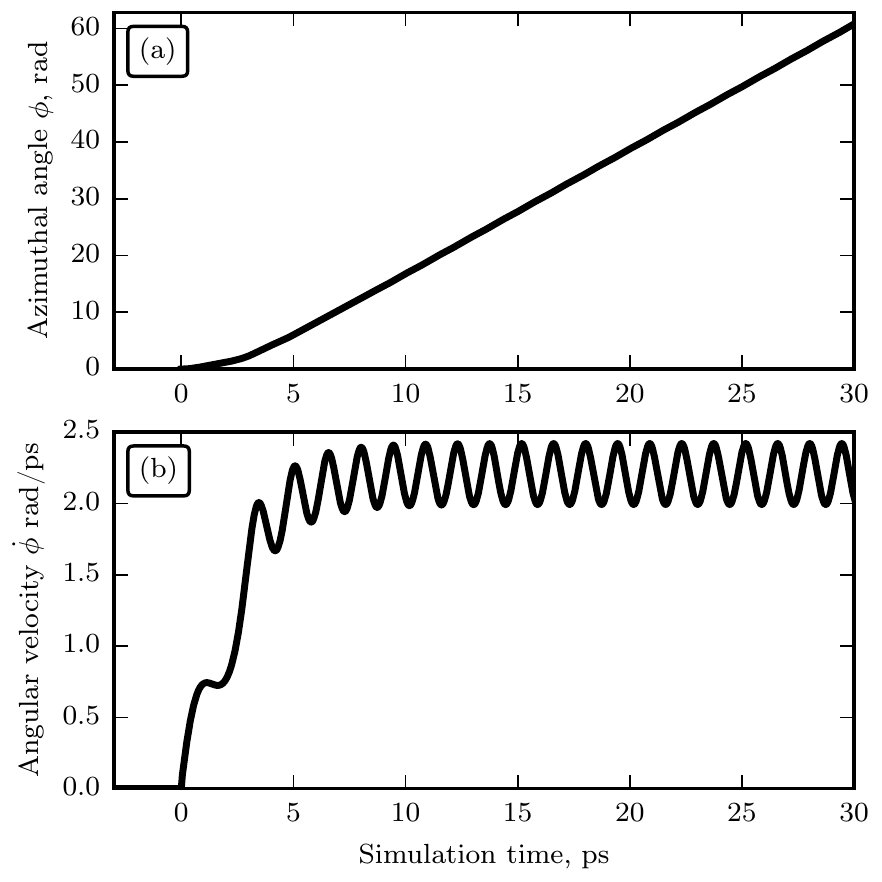}
    \caption{Numerically calculated temporal characteristics of the rotation of the sublattice magnetizations in a bi-axial AFM caused by an abrupt application at $t=0$ of a supercritical spin current ($j>\jthi$ defined by Eq.(4)) polarized along the AFM hard axis: (a)azimuthal angle $\phi$; (b) angular velocity $\dot\phi$ in the AFM easy-plane.
		}\label{fig:mz}
\end{figure}

What is more important, the rotation of the Neel vector, due to the influence of the easy-plane magnetic anisotropy, is \emph{not uniform in time}, $\dot\phi \ne \const$. To illustrate this effect, we solved numerically a system of two coupled Landau-Lifshits equations describing the AFM dynamics for a bilayer \ce{NiO}(\ilu[5]{\nm})/\ce{Pt}(\ilu[20]{\nm}) (see Methods for other calculation parameters). The time dependence of the azimuthal angle $\phi$ after a sudden application of the DC current is plotted in Fig.~\ref{fig:mz}(a). The angle $\phi$ infinitely increases in time as the system makes revolutions around the direction of the spin current polarization~\cite{bib:Cheng:2015}, but its motion \emph{is not uniform in time} due to the action of the easy-plane anisotropy. This non-uniformity is further illustrated by the time dependence of the angular velocity $\dot\phi$, which oscillates in time with THz frequency (see Fig.~\ref{fig:mz}(b)). Thus, the (easy-plane) AFM magnetic anisotropy, existing in the plane perpendicular to the spin current polarization, leads to the generation of an AC output spin-current~\eqref{eq:spin-current-phi}, and turns a current-driven Pt/AFM bi-layer into a potential THz-frequency auto-oscillator.

Another important feature, which is evident from Fig.~\ref{fig:mz}(b), is the existence of a transitional process from the stationary state to a steady oscillatory motion. In contrast to FM materials, the dynamics of AFM is \emph{inertial}~\cite{bib:Kimel:2009, bib:Ivanov:2014}, so that it takes time for the STT to accelerate the sublattice magnetizations, and the magnetic system accumulates some kinetic energy during this transitional process. The inertial nature of the AFM dynamics implies that the AFM oscillator may exhibit a hysteretic behavior, and that two different threshold currents may exist in the AFM oscillator.

\begin{figure}
    \includegraphics[width=0.7\linewidth]{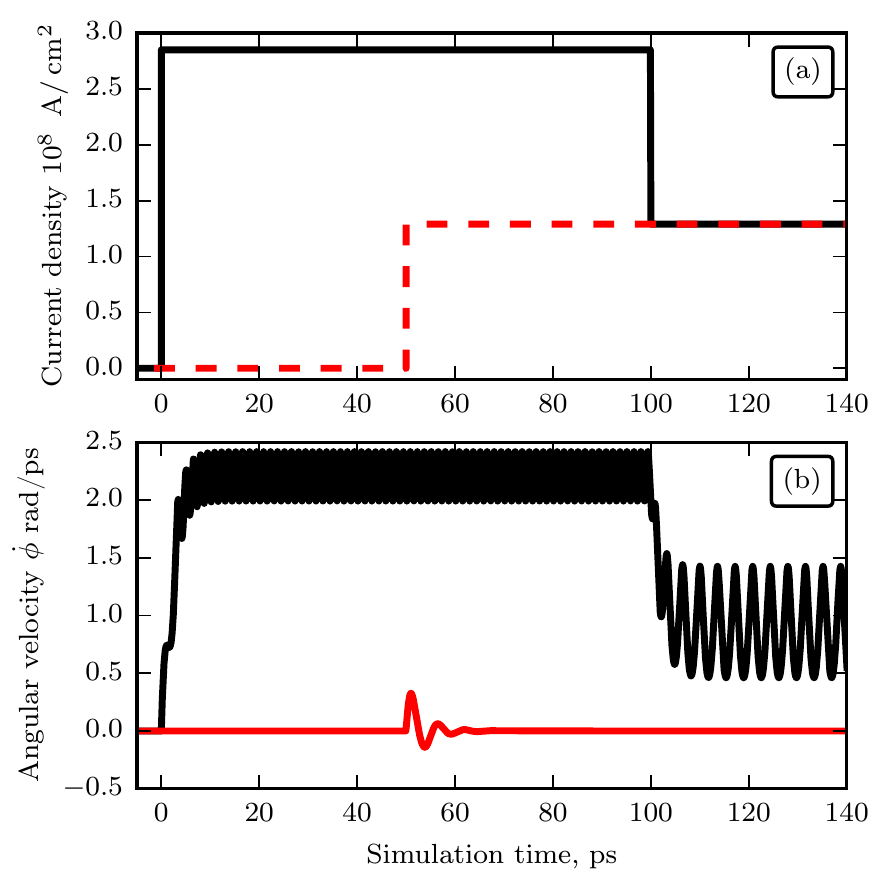}
    \caption{
		Numerically calculated curves illustrating inertial dynamics of an AFM (NiO) under the action of a  DC current step abruptly applied to the adjacent layer of a heavy metal (\ce{Pt}):  (a) DC electric current density in the heavy metal layer, (b) angular velocity of the Neel vector of the AFM layer. Black curves correspond to the case when the initial magnitude to the DC current density , first, is made higher then the ``ignition'' threshold (4), and, then, is lowered at $t=100\picosecond$ to a ``working'' level. Red curves correspond to the case when the magnitude of the DC current density is abruptly increased just to the ``working'' level at $t=50\picosecond$.
		}\label{fig:hyst}
\end{figure}

This process is illustrated by Fig.~\ref{fig:hyst}, which shows the dynamics of the AFM angle $\phi$ after application of electric current. In the first case (black curves) the current, first, overcomes the ``ignition'' threshold $\jthi$, and then is lowered to the ``working'' density, which is below $\jthi$. In the second case (red lines) the current is increased from zero directly to the ``working'' density. In the first case the oscillations continue even after the current was lowered below the ``ignition'' level, while in the second case the same ``working'' current cannot start any oscillations. The existence of the hysteresis in the AFM dynamics allows one to ``ignite'' the generation with a very short (<\ilu[10]{\picosecond}) large-amplitude current pulse, and then to reduce the current density for a continuous oscillator operation. The minimum current needed to \emph{sustain} AFM oscillations (the ``elimination'' threshold) can be found analytically in the limit of a small damping from the condition that the work produced by the STT during one period of rotation equals the energy lost due to the Gilbert damping:
\begin{equation}
    \jthe = \dfrac{2\alpha}{\pi\sigma} \sqrt{\wex \wax}.
    \label{eq:jthe}
\end{equation}

The existence of these two threshold currents is critically important for a practical implementation of the proposed oscillator. For the taken parameters of the bilayer the ignition current density is estimated as $\jthi\approx2\times10^8 \ampere/\squarecentimeter$, which is larger for the already demonstrated FM spin-Hall oscillators~\cite{bib:Demidov:2014}, however, the elimination threshold current density is twice lower $\jthe\approx1.1\times10^8\ampere/\squarecentimeter$, which is \emph{lower}, then the threshold currect densities for the FM spin-Hall oscillators.

Equation~(\ref{eq:phi}) allows one to find approximate analytical solution for supercritical currents $j>\jthi$:
\begin{equation}
    \dot\phi(t) \approx \frac{\wgen}{2} + \dfrac{\wax \wex }
    {4 \sqrt{\alpha^2 \wex^2 + \wgen ^2}}\cos{\wgen t},
    \label{eq:phi_t}
\end{equation}
where $\wgen$ is the AC generation frequency:
\begin{equation}
    \wgen = 2 \dfrac{\sigma j}{\alpha}.
    \label{eq:freq}
\end{equation}
It is clear, that the output AC signal (\ref{eq:spin-current-phi}), proportional to the variable part of the angular velocity $\dot\phi$, is proportional to the anisotropy field $H_e$ in the AFM easy plane, and vanishes for an uniaxial AFM. Also the generation frequency $\wgen$ does not depend on the AFM resonance frequencies, but is determined only by the ratio of STT and the Gilbert damping. Similarly, in JJs, the generation frequency depends on current and not on the Josephson plasma frequency.

\begin{figure}
    \includegraphics[width=0.7\linewidth]{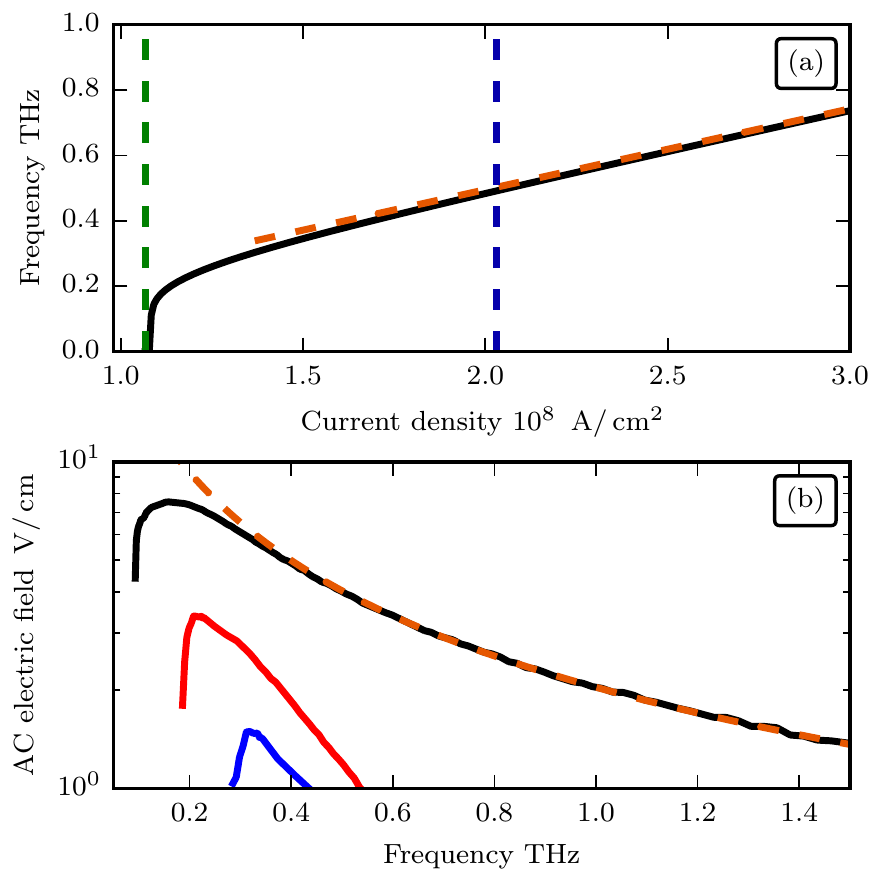}
    \caption{
			(a) Generation frequency $\wgen$ as a function of the DC electric current density. Black solid line shows the results of numerical simulations, orange dashed line is obtained from the approximate formula ~\eqref{eq:freq}, blue and green vertical dashed lines show the ``ignition'' and ``elimination'' threshold current densities, respectively.
		(b) Amplitude of the output AC electric field of the AFM oscillator as a function of frequency. Solid lines show the results of numerical simulations (black line -- amplitude of the fundamental harmonics $\wgen$, red and blue lines -- 2nd and 3rd harmonics, respectively). Dashed orange line corresponds to the analytical formula obtained in \cite{bib:Nakayama:2012} where the approximate expression for the angular velocity~\eqref{eq:phi_t} is used.
		The parameters of the AFM oscillator are given in the Methods.
		}\label{fig:freq}
\end{figure}

Figure~\ref{fig:freq}(a) shows the dependence of the generated frequency $\wgen$ on the current density in the \ce{Pt} layer, while Fig.~\ref{fig:freq}(b) -- dependence of the amplitudes of several harmonics of the output AC electric field as a function of the generated frequency (see Methods for the parameters of the oscillator). The frequency $\wgen$ can be continuously tuned from almost 0 to several THz with current densities $j\sim10^8\ampere/\squarecentimeter$.
The first harmonic of the output AC electric field has a maximum at a relatively low frequency $\simeq$~\ilu[0.1]{\THz}, and, then, slowly decreases, remaining sufficiently large ($>$~\ilu[1]{\volt/\cm}) up to the frequency of \ilu[2]{\THz}. For small frequencies (small currents) the motion of the phase $\phi(t)$ is strongly nonlinear, and the output signal contains multiple higher harmonics, the amplitudes of which decrease rather fast with the increase of the generation frequency.

\section{Discussion}

\label{sec:Discussion}
Above, we demonstrated that a thin layer of a bi-axial AFM (e.g., NiO) driven by a DC spin current flowing in the adjacent Pt layer can work as a THz-frequency auto-oscillator. Below, we compare the proposed antiferromagnetic oscillator (AFMO) to other known sources of coherent microwave or THz-frequency radiation.

A working prototype of a spin-Hall nano-oscillator (SHNO) based on a bi-layer of a FM metal and Pt~\cite{bib:Demidov:2012, bib:Demidov:2014, bib:Duan:2014} had a threshold current density of $j\approx1.3\times10^8\ampere/\squarecentimeter$, which is smaller than the ignition threshold current $\jthi\approx2\times10^8\ampere/\squarecentimeter$, however, this ignition current is needed for only a very short amount of time and can be reduced to lesser current densities above the elimination threshold $\jthe\approx1.1\times10^8\ampere/\squarecentimeter$, making the proposed AFM/Pt configuration promising for the practical implementation.

The frequency of the SHNO is determined mainly by the bias magnetic field and static magnetization of the FM layer,  and was of the order of~\ilu[5-12]{\GHz} for the bias fields ranging between~\ilu[400]{\Oe} and~\ilu[2000]{\Oe}. In the proposed AFMO the generation frequency is determined by the driving electric current, and can be varied from~\ilu[0.1]{\THz} to~\ilu[2.5]{\THz} for experimentally achievable current densities (note, that the current densities of up to $1.1\times10^9\ampere/\squarecentimeter$ have been achieved in \ce{Py/Pt} nanowires~\cite{bib:Duan:2014}).

In Ref.~\cite{bib:Cheng:2016} it was proposed to achieve the THz-frequency generation in NiO/Pt bi-layer via a nonlinear feedback mechanism, which is different from the AFMO generation mechanism described above. In Ref.~\cite{bib:Cheng:2016} the spin-current is polarized along the \emph{easy} axis of the AFM (easy plane anisotropy field in NiO $H_e\approx$~\ilu[628]{\Oe}~\cite{bib:Sievers:1963, bib:Hutchings:1972}), which leads to a rather large threshold current density, because the current-induced STT has to overcome both the large hard axis anisotropy (hard axis anisotropy field in NiO $H_h\approx$~\ilu[15.7]{\kilo\Oe}~\cite{bib:Sievers:1963})) and damping. For the parameters used in this work, we estimate this current density (see Methods) as $j^\text{th}_C\approx5.17\times10^9\ampere/\squarecentimeter$, which is more then $H_h/H_e\approx25$ times larger than the ignition threshold current for the mechanism proposed here,
and, $\approx 50$ times larger then the elimination threshold.
Also, the nonlinear feedback mechanism, which stabilizes the AFM precession around the easy axis in~\cite{bib:Cheng:2016}, works only in a very narrow range of the bias currents and generation frequencies, which severely limits the tunability of the oscillator proposed in ~\cite{bib:Cheng:2016}.

One of the most important characteristics of an oscillator is its output power, which depends on the oscillator's physical dimensions. The devices working at the frequencies of up to several THz must be smaller than the wavelength $\lambda_\text{EM}$ of the electromagnetic radiation at these frequencies. To estimate the output power for the above proposed AFMO we assumed here that the working area of the AFMO has the characteristic dimension $L=10\micrometer\ll\lambda_\text{EM}@1.5\THz\approx200\micrometer$. For such a size the output power can be estimated as $W=E^2 \rho^{-1} L^2 d_{\ce{Pt}}$ (see Methods for parameters). Thus, for the proposed AFMO, the output voltage varies from \ilu[6]{\millivolt} to \ilu[1]{\millivolt} in the frequency range from~\ilu[0.1]{\THz} to~\ilu[2.0]{\THz}, which gives the power range from \ilu[1.5]{\micro\watt} to \ilu[40]{\nano\watt}.

The known oscillators working in the THz-frequency range are the current-biased Josephson junction oscillators working at cryogenic temperatures and having a similar power range~\cite{bib:Darula:1999}, but a much smaller output voltage (\ilu[16]{\micro\volt} in Ref.~\cite{bib:Ulrich:1973}). The amplitude of the output voltage (and, consequently, the output power)of the above proposed AFMO working at room temperature depends on the magnitude of the in-plane anisotropy in the AFM layer (see~\eqref{eq:phi_t}). To increase the output voltage at higher generation frequencies one can use AFM materials with a stronger easy-plane  anisotropy, but this, obviously, will lead to  a corresponding increase in the ``ignition'' threshold current density.

In conclusion, we demonstrated theoretically, that a pure spin current in the AFM layer induced by a DC driving electric current flowing in the adjacent \ce{Pt} layer can excite THz-frequency oscillations. We showed, that in the case of a \ce{NiO}(\ilu[5]{\nm})-\ce{Pt}(\ilu[20]{\nm}) bi-layer it is possible to achieve the generation of \ilu[0.1-2.0]{\THz} signals with reasonable current densities that were previously achieved in FM SHNO of a similar geometry.  The estimated  AC voltage of the proposed AFMO, which can be  picked up using the  inverse spin-Hall effect in the Pt layer can exceed \ilu[1]{\volt/\cm} for the generation frequencies of up to \ilu[2]{\THz}.

\section{Methods}
\label{sec:Methods}
\subsection{Spin-transfer torque}
In this work $\tau$ is the amplitude of the STT expressed in the units of frequency ~\cite{bib:Tserkovnyak:2005}:
\begin{equation}
\tau=j \theta_{SH} g_r \frac{e \gamma \lambda \rho}{2 \pi M_s d_{\ce{AFM}}} \tanh \frac{d_{\ce{Pt}}}{2 \lambda} = \sigma j,
\label{eq:tau}
\end{equation}
where $j$ is the density of the driving DC electric current in the \ce{Pt} layer,
$g_r$ is the spin-mixing conductance at the \ce{Pt-AFM} interface, $\lambda$ is the spin-diffusion length in the \ce{Pt}, $\rho$ is the \ce{Pt} electric resistivity, $\Ms$ is the saturation magnetization of one of the AFM sublattices, and $d_{\ce{AFM}}$ and $d_{\ce{Pt}}$ are the thicknesses of the \ce{AFM} and \ce{Pt} layers, respectively.

\subsection{Coupled Landau-Lifshitz equations}

Dynamics of a thin antiferromagnetic film can  be numerically simulated using two coupled Landau-Lifshitz-Gilbert-Slonczewski equations~\cite{bib:Gurevich:1996, bib:Gomonay:2014, bib:Berger:1996, bib:Slonczewski:1996} for two magnetic sublattices $\vec{M}_1$ and $\vec{M}_2$ of the AFM:
\begin{subequations}
\begin{eqnarray}
d\mathbf{M}_1/d t=\gamma[\mathbf{H}_1 \times \mathbf{M}_1]+\dfrac{\alpha}{\Ms}[\mathbf{M}_1 \times d \mathbf{M}_1/ d t]+\dfrac{\tau}{\Ms}[\mathbf{M}_1 \times [\mathbf{M}_1 \times \vec{p}]], \\
d\mathbf{M}_2/d t=\gamma[\mathbf{H}_2 \times \mathbf{M}_2]+\dfrac{\alpha}{\Ms}[\mathbf{M}_2 \times d \mathbf{M}_2/ d t]+\dfrac{\tau}{\Ms}[\mathbf{M}_2 \times [\mathbf{M}_2 \times \vec{p}]],
\end{eqnarray}
\label{eq:LLs}
\end{subequations}
where $\alpha$ is the effective Gilbert damping parameter, $\tau$ is the STT expressed in the frequency units (see equation~\eqref{eq:tau}), $\vec{p}$ is a unit vector along the spin current polarization, and $\mathbf{H}_1$ and $\mathbf{H}_2$ are the effective magnetic fields acting on the sublattices $\mathbf{M}_1$ and $\mathbf{M}_2$, respectively:
\begin{subequations}
\begin{eqnarray}
\mathbf{H}_1=\dfrac{1}{\Ms}\left[-\dfrac{1}{2}H_{\text{ex}} \mathbf{M}_2+ H_h \mathbf{n}_h (\mathbf{n}_h \cdot  \mathbf{M}_1) - H_e \mathbf{n}_{e}  (\mathbf{n}_{e} \cdot  \mathbf{M}_1)\right],\\
\mathbf{H}_2=\dfrac{1}{\Ms}\left[-\dfrac{1}{2}H_{\text{ex}} \mathbf{M}_1+ H_h \mathbf{n}_{h} (\mathbf{n}_{h} \cdot  \mathbf{M}_2) - H_e \mathbf{n}_{e} (\mathbf{n}_{e} \cdot  \mathbf{M}_2)\right].
\end{eqnarray}
\end{subequations}
Here $H_{\text{ex}}$ is the exchange field, $H_e$ and $H_h$ are the easy-plane and hard-axis anisotropy fields, respectively, and $\vec{n}_e$ and $\vec{n}_h$ are the unit vectors along the hard and easy axes.

\subsection{Antiferromagnetic dynamics}
To study the dynamics of AFM analytically we use the standard $\sigma$-model~\cite{bib:Andreev:1980,bib:Ivanov:2014}.
The coupled Landau-Lifshits equations~\eqref{eq:LLs} can be rewritten in terms of the $\vl$ and $\vm$. Assuming that $|\vm|\ll|\vl|$, which is valid when the exchange field $H_{\text{ex}}$ is larger than any other field acting in the AFM, one can consider $\vm$ as a slave variable:
\begin{equation}
    \vm = \dfrac{1}{\gamma H_\text{ex}} \vl\cross\dot\vl.
\end{equation}
In this approximation the dynamics of $\vl$ is governed by one second-order vectorial differential equation:
\begin{equation}
    \vl\cross\left[\dfrac{1}{\wex}\ddot\vl +\alpha \dot\vl + \op{\Omega}\cdot\vl + \tau\vec{p}\cross\vl \right] = 0,
    \label{eq:vec_l}
\end{equation}
with an additional constraint $|\vl|=1$. Here $\op\Omega = \gamma (H_e \vec{n}_e\otimes\vec{n}_e - H_h\vec{n}_h\otimes\vec{n}_h)$, and the symbol $\otimes$ denotes the direct vector product. Equation~\eqref{eq:vec_l} is effectively two-dimensional, so we can rewrite it in a spherical coordinate system. To simplify the analytical derivation we assume that $\z=\vec{p}$ (i.e., the spin current is polarized along the $\z$-axis), $\vec{n}_e = \x$, and $\vec{n}_h=\z$:
\begin{eqnarray}
\frac{\dot \theta \dot \phi}{\wex}\sin2\theta+ \sin^2 \theta \left[\dfrac{\ddot \phi}{\wex}+\alpha \dot \phi - \frac{\wax}{2} \sin 2 \phi + \tau\right]=0. \label{motion-1}\\
\frac{\ddot \theta}{\wex}+\alpha \dot \theta - \frac{\sin 2 \theta}{2} \left[\frac{\dot \phi^2}{\wex}-\wax \cos^2 \phi + \waz \right]=0, \label{motion-2}
\label{motion}
\end{eqnarray}
where $\omega_e = \gamma H_e$ and $\omega_h=\gamma H_h$. The ground state of the AFM corresponds to $\theta=\pi/2$ and $\phi=0$. The solution $\theta=\pi/2$ (i.e., vector $\vl$ rotates in the $xy$-plane) is stable for the considered geometry~\cite{bib:Gomonay:2010,bib:Cheng:2016} and automatically satisfies the equation~\eqref{motion-2}. Using $\theta=\pi/2$ in equation~\eqref{motion-1} it is possible to obtain a single equation~\eqref{eq:phi} for the azimuthal angle $\phi$.

\subsection{Parameters of the system}
In all the numerical simulations and estimations reported here we considered a Pt/NiO bilayer with the following parameters:
thickness of the \ce{NiO} and \ce{Pt} layers $d_{\ce{AFM}}= 5\nanometer$ and $d_{\ce{Pt}}= 20\nanometer$,
spin diffusion length in \ce{Pt} $\lambda=7.3\nanometer$~\cite{bib:Wang:2014},
electrical resistivity in \ce{Pt} $\rho=4.8\times 10^{-7}\ohmm$~\cite{bib:Wang:2014},
spin-mixing conductance at the \ce{Pt-NiO} interface  $g_r=6.9\times 10^{18} \meter\unitsuperscript{-2}$~\cite{bib:Cheng:2014},
magnetic saturation of one \ce{NiO} sublattice $\Ms=351\kA/\meter$~\cite{bib:Hutchings:1972},
spin-Hall angle in Pt $\theta_{SH}=0.1$~\cite{bib:Wang:2014},
effective Gilbert damping is  $\alpha=3.5\times 10^{-3}$ (see below),
exchange frequency $\wex=2\pi\times 27.5\THz$,
$\gamma H_e = \wax=2\pi \times 1.75\GHz$ and $\gamma H_h = \waz = 2\pi\times43.9\GHz$~\cite{bib:Sievers:1963}.
For the chosen parameters the coefficient $\sigma$ in~\eqref{eq:tau} is $\sigma/(2\pi)=4.32\times 10^{-4} \hertz\squaremeter/\ampere$.

\subsection{Output electric field}
The rotation of the vector $\vl$ in the AFM layer induces a spin-current into the adjacent\ce{Pt} layer, which, in turn, creates an electric field in the \ce{Pt} layer via the inverse spin-Hall effect (ISHE). This AC electric field serves as the output signal of the AFMO. The ISHE electric field is calculated using the following analytic expression~\cite{bib:Nakayama:2012}:
\begin{equation}
    E = \theta_\text{SH} \dfrac{g_re\lambda\rho}{2\pi}\dfrac{1}{d_{\ce{Pt}}}\tanh\left(\dfrac{d_{\ce{Pt}}}{2\lambda}\right) \dot\phi = \kappa \dot\phi.
\end{equation}
For the chosen parameters of the AFM oscillator the parameter $\kappa$ is equal to $\kappa \approx 1.35\times10^{-9} \volt/\meter\cdot(\rad/\second)^{-1}$.

\subsection{Effective Gilbert damping}
The intrinsic Gilbert damping constant $\alpha_0$ for \ce{NiO} can be calculated from the experimentally measured linewidth
$\Delta\omega_{\text{AFMR}}/(2\pi) = 18\GHz$
of the AFM resonance~\cite{bib:Sievers:1963, bib:Nishitani:2010}.
The linewidth $\Delta \omega_{\text{AFMR}}$ is related to $\alpha_{0}$ by $\Delta\omega_{\text{AFMR}} =\alpha_0\omega_{\text{ex}}$, where $\omega_{\text{ex}} = \gamma H_{\text{ex}} = 2 \pi \cdot 27.5 \THz$ is the exchange frequency~\cite{bib:Hutchings:1972}. One can see, that the intrinsic Gilbert damping in NiO is rather small: $\alpha_0\simeq 6 \cdot 10^{-4}$. However, the spin pumping from NiO to Pt layer can be described as an additional damping mechanism for the spin dynamics in the AFM~\cite{bib:Nakayama:2012, bib:Tserkovnyak:2005}, and the total effective damping constant can be written as:
\begin{equation}
	\alpha=\alpha_0 + g_r \frac{\gamma \hbar}{4 \pi M_s d_{\text{AFM}}}.
\end{equation}
For the chosen parameters of the AFMO the damping parameter is $\alpha = 3.5\times10^{-3}$. Thus, the effective damping in thin Pt/NiO bi-layers is dominated by the spin-pumping mechanism, and strongly depends on the NiO thickness.

\subsection{Threshold current in the case when the spin current is polarized along the easy axis}
Cheng \emph{et. al.} estimated the threshold STT needed to start the oscillations in the case when the spin current is polarized along the easy axis as (see eq.~(3) in Ref.~\cite{bib:Cheng:2016}):
\begin{equation}
    \tau^\text{th}_C = \sqrt{\waz^2/4 + \alpha^2 (2\wax + \waz)\wex}
\end{equation}
For the parameters of our AFMO, the threshold electric current for the generation mechanism described in Ref.~\cite{bib:Cheng:2016}can be calculated as:
\begin{equation}
    j^\text{th}_C = \tau^\text{th}_C/\sigma = 5.17\times 10^9 \ampere/\squarecentimeter.
\end{equation}

\section{Acknowledgements}
This work was supported in part by Grant No. ECCS-
1305586 from the National Science Foundation of the USA,
by the contract from the US Army TARDEC, RDECOM,
and by the Center for NanoFerroic Devices (CNFD) and the
Nanoelectronics Research Initiative (NRI).
BAI was partly supported by the National Academy of
Sciences of Ukraine, project No~1/16-N.

\section{Author contributions}
RH, IL, and VT developed an analytical model of AFM excitation under the action of STT. RH performed the numerical simulations. AS and BAI formulated the problem and supervised the work. All the authors contributed to the text.

\section{Competing financial interests}
The authors declare an absence of competing financial interests.

\bibliography{afm_gen}

\end{document}